# Russian Cyber Onslaught was Blunted by Ukrainian Cyber Resilience, not Merely Security

*Alexander Kott, Independent Consultant*

*George (Yegor) Dubynskyi, Ministry of Digital Transformation of Ukraine and G.E. Pukhov Institute for Modeling in Energy Engineering, Ukraine*

*Andrii Paziuk, National Aviation University, Ukraine*

*Stephanie E. Galaitsi, US Army Engineer Research and Development Center*

*Benjamin D. Trump, US Army Engineer Research and Development Center*

*Igor Linkov, US Army Engineer Research and Development Center*

Abstract: Russian cyberattacks on Ukraine largely failed to produce meaningful outcomes not merely due to robust Ukrainian cyber defenses but were instead primarily a result of Ukraine's effective cyber resilience.

Before the full-scale February 2022 Russian invasion of Ukraine, global military and political analysts projected a swift Russian victory through overwhelming force and superior capabilities. It was anticipated that Russian military prowess, characterized by formidable, armored units and advanced missile technology, would quickly overwhelm Ukrainian defenses, while Russian cyber operations would destroy Ukraine's communication infrastructure, disrupt command and control systems, neutralize air defenses, and cripple critical infrastructure such as the electric grid. This cyber component was expected to be particularly devastating, potentially severing the Ukrainian government's internal communications and isolating it from the populace.

Contrary to these predictions, the invasion unfolded differently. Ukrainian forces not only halted the advancement of Russian armored columns but also inflicted significant casualties and in some places compelled their retreat. While Russia launched coordinated cyberattacks alongside physical assaults, these efforts to disrupt critical Ukrainian cyber infrastructure were never successful enough to discernibly benefit the Russian campaign. Notably, the primary telecommunications networks remained operational, though certain cyberattacks did disrupt Ukrainian infrastructure, such as the disabling or defacing of many websites of the Ukrainian government, media, and commercial organizations. Granted, some degree of success in Russia's cyberattacks did occur. The attacks were at least partly coordinated with missile attacks and with advances of ground forces. Some attacks successfully disabled or defaced many websites of the Ukrainian government, media, and commercial organizations, and other attacks placed pro-Russian propaganda and calls for surrender on many Internet resources [1]. Ultimately, few if any of Russia's presumed objectives were achieved; few Ukrainians paid attention to transparent Russian propaganda on temporarily captured websites, while critical functions of the Ukrainian cyber infrastructure remained intact. It was these critical functions that carried the "We Are Here" speech by President Volodymyr Zelenskyi to millions of Ukrainians, calling them to arms shortly after the invasion began [2].

Throughout the conflict, the intensity and frequency of Russian cyberattacks escalated, yet their effectiveness remained minimal. Several theories have been proposed to explain this apparent shortfall in Russian cyber capabilities. One suggestion is that Russia reserved its most formidable cyber weapons for potential future conflicts with NATO or the US, or for a future decisive moment in the invasion [3]. Another theory argues that cyber warfare is inherently ineffective at crippling military capacities and that Russia instead shifted focus to intelligence gathering and information operations [3–5]. A third perspective challenges the assumption of Russia's potent cyber arsenal, proposing that Russia's actual cyber capabilities were insufficient for significant impacts [4].

Despite various speculations, one conclusion is evident: Russian cyberattacks largely failed to produce meaningful outcomes. In this article we discuss how such failures were not merely due to robust Ukrainian cyber defenses but were instead primarily a result of Ukraine's effective cyber resilience. *Cyber resilience* denotes recovery from disruption, where recovery can vary by both magnitude and speed. An efficient recovery in one sector can effectively insulate other sectors from secondary effects. This document explores the distinctions and interplay between cyber security and cyber resilience within the context of the Russo-Ukrainian conflict to demonstrate that 'security' and 'resilience' must be framed as complementary yet separate concepts that influence the operational capability of various infrastructure systems performing amidst challenging and violent conditions of war.

## Cyber resilience vs Cyber security

The concepts of resilience and security, particularly when prefixed by "cyber," are often mistakenly used interchangeably. Definitionally, security is defined as "the state of being free from danger or threat" [6]. This definition aligns closely with that of risk, which is described as "a situation involving exposure to danger (threat)" [6]. Therefore, *cyber security* focuses on preventing and reducing exposure to these risks. In contrast, *resilience* is defined as "the capacity to recover quickly from difficulties" [6], underscoring that it applies to situations only after a risk has materialized into an actual disruption. The distinction between the two is critical: cyber security is about risk avoidance—preventing threats from affecting the system—while cyber resilience deals with response and recovery, ensuring continuity and restoration of operations after security measures have been bypassed or compromised.

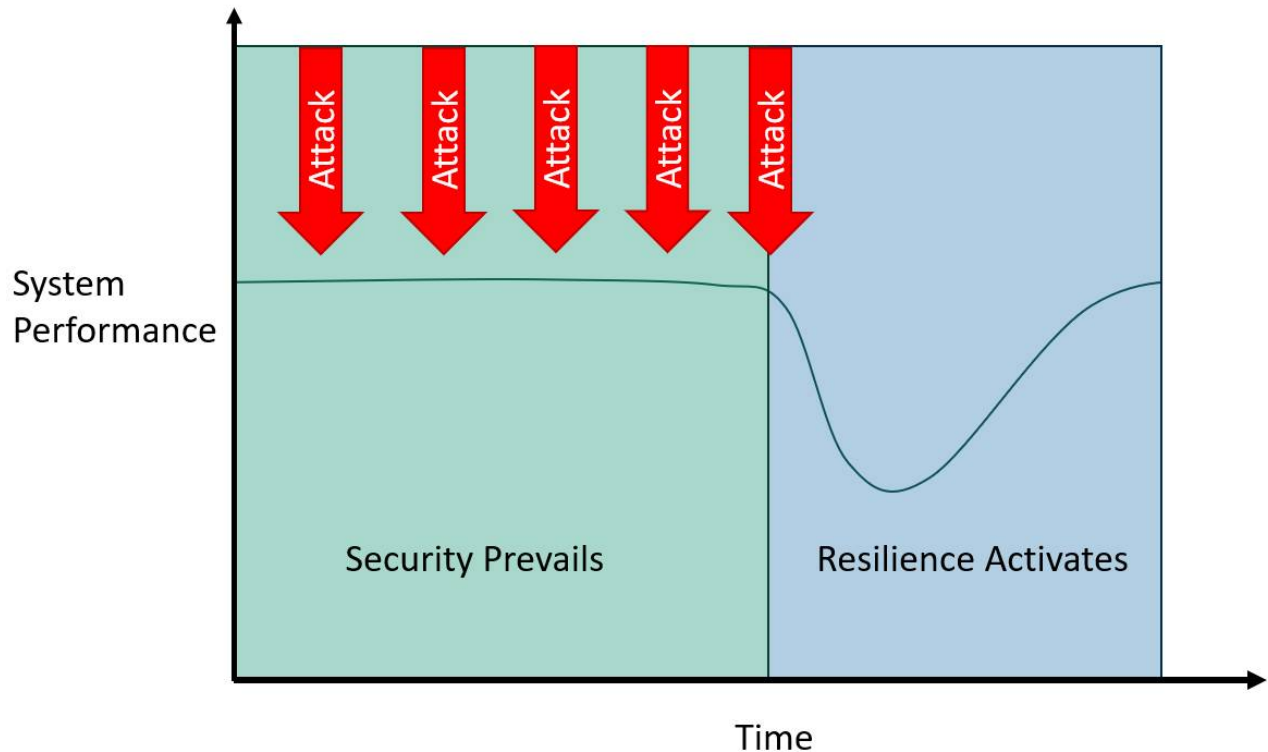

The United States National Institute of Standards and Technology (NIST) distinguishes between cyber security and cyber resilience as follows. NIST defines cyber security as the "prevention of damage to, protection of, and restoration of computers, electronic communications systems, electronic communications services, wire communication, and electronic communication, including information contained therein, to ensure its availability, integrity, authentication, confidentiality, and nonrepudiation"[7]. Conversely, cyber resilience is defined as "the ability to anticipate, withstand, recover from, and adapt to adverse conditions, stresses, attacks, or compromises on systems that use or are enabled by cyber resources" [8]. Cyber resilience is designed to sustain mission-critical operations even when the cyber environment is under threat, making it indispensable in contested environments.

Both concepts embody elements of robustness—cyber security through "prevention" and cyber resilience through "withstanding." They each also reference resilience: cyber security through "restoration" and cyber resilience through "recover[y]." Thus, these concepts need further differentiation. Cyber security aims to shield the system from damage, whereas cyber resilience aims to restore normalcy after damage occurs. It is important to understand when to apply each of these disciplines.. Employing cyber security strategies is ineffective once a breach has occurred, just as relying solely on resilience strategies is insufficient for preventing initial compromises. This differentiation is particularly important in environments like the ongoing Russo-Ukrainian conflict, where the cyber domain is heavily contested, and the likelihood of compromises is high. In such scenarios, the focus shifts toward cyber resilience due to the increased probability of disruptions needing effective recovery strategies.

The overlap and misapplication of these terms not only confuse their practical implementations but can lead to costly strategic errors. Cyber resilience becomes increasingly important to performance in scenarios where threats are difficult to characterize and can affect a multitude of potential vulnerabilities across a vast complex infrastructure network. Such circumstances necessitate a dynamic and flexible response capability to maintain system integrity and operational continuity. This paradigm has been vividly demonstrated in the context of Ukraine's response to Russian cyber aggression.

# Paradigms of Cyber Resilience

Cyber resilience utilizes an array of strategies to maintain system functionality and optimize recovery processes under cyberattack conditions. These strategies encompass various paradigms that collectively fortify systems against a wide range of threats [9]. Such strategies can be categorized as follows:

- The **Discarding and Replacement** paradigm involves the elimination of system elements that are irreparably damaged or compromised. These components are either replaced with intact ones or removed entirely to allow the system to function at a diminished capacity. This process ensures that critical functionalities continue, albeit with potential limitations, by purging severely affected elements that could hinder system operations.

- In the **Deflection and Absorption** approach, the effects of cyber threats are redirected from primary targets to less critical system elements. This deflection helps to absorb the impact of the attack within a less critical system component while main functionalities remain operational, allowing for targeted fortifications and repairs.

- The **Reconfiguration, Resource Re-allocation, and Relocation** paradigm involves adaptive modifications to the system's setup, which might include adjusting the functionalities of certain components or altering the distribution of resources to prioritize critical operations. This could also extend to the physical or digital relocation of assets to secure locations. These actions are crucial for maintaining the availability and integrity of essential services and protecting sensitive data under adverse conditions.

- **Dynamic Generation of Information, Services, and Connections** strategy ensures the continuous update and regeneration of system services and connections to prevent stagnation in any single configuration, which could be exploited by attackers. By frequently refreshing its operational state and managing the lifecycle of its connections, the system maintains a moving target posture, enhancing its defense against persistent or evolving cyber threats.

Together, these paradigms have been instrumental in Ukraine's defense against Russian cyber operations, showcasing their effectiveness in a real-world application. Each paradigm not only supports resilience in its unique way but also complements the others to create a robust, multi-layered defense architecture. The ensuing sections will provide detailed examples of how these strategies have been deployed to safeguard Ukraine's cyber infrastructure throughout ongoing hostilities.

## Discarding and Replacement: Viasat and Starlink

One mechanism of resilience – discarding and replacement – starts by identifying the compromised components or subsystems of the overall system and, if impossible to fix, discarding them. Then, replace them with a different one, not yet compromised and hopefully more difficult to compromise. The transition from Viasat to Starlink, which began almost immediately after the initial attack, serves as an example.

On February 24, 2022, about an hour before Russian ground forces started to enter the territory of Ukraine, Russian hackers working for the intelligence organization GRU initiated a denial-of-service attack that forced many Viasat models offline. This was followed by an attack on a ground-based network that allowed malware to enter a management segment of the communications satellite K-SAT network. The malware AcidRain was involved in the process. From the management segment, the cyber actors issued commands to thousands of modems; the commands overwrote data in the flash memory of the modem and made them incapable of performing key functions.

Although some of Viasat's equipment was quickly restored, many thousands of replacement modems had to be brought into Ukraine and re-installed. The Ukraine decision-makers elected to essentially follow the "discard and replace" approach – discard Viasat and replace it with Starlink. Only four days after the Viasat hack, Starlink terminals began to arrive in Ukraine, eventually reaching numbers in tens of thousands.

While the attack on Viasat did cause a significant disruption of Ukraine government communications in the early weeks of the Russian invasion, there is no confirmation that specifically military channels were affected. More broadly, there is no evidence that the attack yielded any tangible military benefits to the Russian invasion. Furthermore, by replacing the Viasat systems with Starlink, Ukraine's military and government not only restored the pre-invasion capacity but improved the military-specific functions of their satellite communications. Starlink has reportedly enabled the control of Ukrainian drones, both aerial and maritime, in ways that the Viasat system might be ill-suited to support [4,10].

## Deflection and Absorption: Highly Decentralized Networks

The deflection and absorption strategy was effectively utilized due to Ukraine's decentralized telecommunication network architecture. A successful attack on one of the many operators could readily absorbed by shifting the load, and thereby functionality, to unaffected operators. This framework allowed the immediate redirection of data flows away from compromised nodes to intact ones, minimizing service disruptions and maintaining network integrity. The diverse and decentralized patchwork of Ukraine's telecommunication providers – hundreds of separate and geographically distant mobile and Internet providers – enabled a 'resilience-by-design' approach, allowing for extensive redundancy and adaptive capacity that supports deflections. This network's design inherently lacks centralized chokepoints, which disperses cybersecurity risks and prevents widespread system failures from any single cyberattack [4].

To ensure uninterrupted service for Ukrainian users, telecommunications operators extensively collaborated to implement universal national roaming. This arrangement allowed subscribers of an operator compromised by a cyberattack to seamlessly switch to networks of other providers that remained unaffected [11]. Despite numerous and continuous adversarial cyberattacks, which caused significant Internet disruptions, these efforts did not critically damage people's ability to access Ukraine's telecommunications infrastructure. As a result, the overall capacity of the network sustained enough functionality to support both civilian and military communications throughout the conflict, minimizing the impact on Ukraine's war efforts [12].

# Reconfiguration, Resource Re-allocation, and Relocation: Data in Western Clouds

A significant application of the reconfiguration and resource re-allocation paradigm is observed in Ukraine's strategic move to transfer essential data assets to cloud servers situated outside its geographical boundaries, predominantly across various secure locations in Europe. This strategic migration primarily aimed to mitigate the risk of Russian interference, which was notably less feasible externally compared to local servers within Ukraine.

In the lead-up to the Russian invasion, the Ukrainian government enacted legislation permitting the transfer of critical governmental and private sector data to foreign cloud systems. These cloud systems, characterized by their decentralized nature and robust security measures—including extensive geographical distribution, data redundancy, and strong encryption protocols—offer enhanced protection against both physical and cyber threats, ensuring data integrity and availability even under dire circumstances.

Aggressive and continuous cyber and physical threats prompted a swift and extensive data migration. Within just the first few months following the outbreak of the war, numerous entities spanning government, academia, and the commercial sector had successfully relocated their digital assets to the cloud [13–16]. This proactive shift not only safeguarded sensitive information but also provided a resilient foundation to rapidly restore and reconfigure systems as needed. For instance, telecommunications providers leveraged cloud infrastructure to maintain operational continuity amid cyberattacks. By regularly updating their backups on secure cloud servers, these companies could quickly recover lost data and restore critical functionalities following an attack. This capability was dramatically demonstrated in March 2022 when Ukrtelecom, after suffering a debilitating cyberattack that eliminated 87% of its network capability, managed to restore full connectivity within merely 15 hours [17–19].

The persistence of Russian cyber offensives was further evidenced in December 2023, nearly two years post-Ukrtelecom incident, when another major provider, Kyivstar, became the target of a sophisticated cyberattack orchestrated by the GRU's Sandstorm unit. This attack devastated thousands of virtual servers, significantly disrupting services. Despite the severity of the attack and the initial failure of security measures, Kyivstar's comprehensive pre-planned resilience strategies enabled a rapid recovery, fully reinstating operational capabilities within a week [18–22].

The consistent reliance on cloud technologies to enable quick data recovery and system reconfiguration exemplifies the practical benefits of the "Reconfiguration, Resource Re-Allocation, and Relocation" paradigm. By decentralizing their data storage and management, Ukrainian entities not only enhanced their defensive posture against the ongoing cyber warfare but also maintained essential services and communications, which proved critical in sustaining national defense operations and civilian welfare during the conflict.

This decentralized, highly entrepreneurial infrastructure also demands a broadly competent workforce that had diverse skills to repair rapidly a variety of telecommunication systems. The robust, collaborative IT workforce across Ukraine rapidly mitigated and repaired damage when attacks occurred. This agility was critical in maintaining operational continuity and was supported by the implementation of universal national roaming, which ensured that users of a compromised operator could seamlessly switch to unaffected networks [23].

## Dynamic generation of information, services, and connections: Link to the U.S. European Command

The dynamic generation of information, services, and connections is crucial in maintaining secure communication during cyber conflicts. This paradigm involves the continuous creation and updating of resources to ensure they remain secure from adversary compromises. The utilization of cloud backups, as previously discussed, is a prime example of how critical data can be dynamically and securely managed. In the realm of secure communications, particularly notable is the establishment of a direct, highly secure communication line between the Ukrainian military and the U.S. European Command right before the Russian invasion in 2022. This line was specifically designed to be impervious to detection and interference by hostile forces, distinguishing it from pre-existing communication channels between Ukraine and its Western allies.

This newly established link offered a reliable, secure channel unknown to Ukrainian adversaries, significantly reducing the likelihood of compromise. Throughout the escalation of the conflict, this dedicated line enabled consistent and secure communications between the Ukrainian military and the U.S. European Command, playing a pivotal role in coordination and strategic planning [24]. This strategy of dynamically generating and securing communication links underscores the adaptability and foresight within Ukrainian cyber defense measures, enhancing their overall resilience against Russian cyber operations.

While emphasizing the crucial role of resilience, we must also acknowledge the significance of security in cybersecurity frameworks. The conflict between Russia and Ukraine has highlighted several instances where Ukrainian cyber defenses successfully thwarted attacks that could have led to severe consequences. For example, in April 2022, Ukraine's cybersecurity forces detected and neutralized malware aimed at electrical substations that, if successful, could have disrupted power for approximately two million people [25,26].

However, achieving flawless security is rarely possible, particularly against sophisticated and well-equipped adversaries. This reality underscores the necessity of cyber resilience — the capacity of a system to sustain damages yet quickly recover to maintain operational continuity. Throughout the ongoing conflict, Ukraine has been subjected to numerous cyberattacks that, despite causing disruptions, did not decisively affect the overall war effort, primarily due to effective and rapid recovery processes [27,28].

Ideally, security measures would prevent any damage from occurring in the first place, embodying the principle of prevention over treatment. Yet, in practice, when security measures are compromised, resilience strategies become indispensable to ensure that systems can quickly return to normal operations and mitigate the impact of attacks. Thus, while security aims to shield against potential threats, resilience prepares the system to cope with and recover from any breaches swiftly, providing a robust approach to recovering critical systems' functionality under duress. This balanced focus on both prevention and recovery is essential for constructing a comprehensive cyber defense strategy.

# Concluding Remarks

To date, the effectiveness of Russian cyber warfare against Ukraine has notably failed to impact the war's overall trajectory. Despite attempts at undermining international support for Ukraine through information campaigns, the internal stability and operational continuity within Ukraine have remained largely intact. The resilience and robust cybersecurity practices adopted by Ukraine have played critical roles in thwarting Russian cyber attacks.

Time and again, Ukrainian cyber defenses have successfully repelled attacks on critical infrastructure, preventing any substantial damage. On occasions where breaches did occur, the rapid response and recovery processes implemented by Ukrainian forces minimized potential impacts, showcasing a high degree of cyber resilience that prevented the kind of widespread disruption that could cripple a less prepared nation [29].

Moreover, as the frequency of attacks increased, the effectiveness of Russian strategies did not follow suit, leading to a declining success rate [30]. This outcome not only highlights the futility of the resources expended by Russia but also suggests a significant exposure and weakening of its own cyber capabilities. By March 2022, Russia had become one of the most frequently targeted nations in cyberspace, often by independent vigilante groups, which unexpectedly also brought Western attention to the vulnerabilities within Russia's cyber defenses [31].

Ukraine's handling of these cyber threats underscores the economic logic behind investing in both cyber resilience and cybersecurity. The balance between these investments should be strategically aligned with the anticipated scale and impact of potential disruptions. Proactive security investments are justified where the risks of immediate and severe consequences are high, whereas uncertain threats might be better addressed through enhanced resilience measures that address recovery after unprevented disruptions. Regardless of the specific threat landscape, maintaining a dual focus on resilience and robust security measures is essential to safeguard continuous operation and service provision, adapting to the unpredictable nature of cyber adversaries. This integrated approach not only preserves critical infrastructures but also fortifies national defense capabilities against a spectrum of cyber threats.

ALEXANDER KOTT is a science and technology consultant and former Chief Scientist of the U.S. Army Research Laboratory. Contact him at alexkott@yahoo.com.

GEORGE (YEGOR) DUBYNSKYI is the Deputy Minister for Cybersecurity, Ministry of Digital Transformation of Ukraine and PhD student at G.E. Pukhov Institute for Modeling in Energy Engineering. Contact him at dubynskyi@thedigital.gov.ua.

ANDRII PAZIUK is a Professor at the National Aviation University, Ukraine. Contact him at inet.media.law@gmail.com.

STEPHANIE E. GALAITSI is a research scientist with the U.S. Army Corps of Engineers' Engineer Research and Development Center. Contact her at stephanie.e.galaitsi@usace.army.mil.

BENJAMIN D. TRUMP is a Senior Research Social Scientist at U.S. Army Engineer Research and Development Center. Contact him at benjamin.d.trump@usace.army.mil.

IGOR LINKOV is the Senior Scientific and Technical Manager at U.S. Army Engineer Research and Development Center, US Army Corps of Engineers. Contact him at Igor.Linkov@usace.army.mil.

**Acknowledgements** This paper is based upon work supported by the US Department of Defense Operational Resilience International Cooperation Program (DORIC). The views and opinions expressed in this article are those of the individual authors and not those of the U.S. Army, Ministry of Digital Transformation Ukraine or other sponsor organizations.